\documentstyle[multicol,aps,epsf]{revtex}
\begin{document}
\draft
\title{Simulations of liquid crystal hydrodynamics}
\author{Colin Denniston$^1$, E. Orlandini$^2$, and J.M. Yeomans$^1$}
\address{$^{1}$ Dept. of Physics, Theoretical Physics,
University of Oxford, 1 Keble Road, Oxford OX1 3NP}
\address{$^{2}$ INFM-Dipartimento di Fisica, Universit\`a di
  Padova, 1-35131 Padova,Italy}

\date{\today}
\maketitle

\begin{abstract}
We present a lattice Boltzmann algorithm for liquid crystal
hydrodynamics.  The coupling between the tensor order parameter and
the flow is treated consistently allowing investigation of a wide
range of non-Newtonian flow behavior.  We present results for the
effect of hydrodynamics on defect coalescence; of the appearance of
the log-rolling and kayaking states in Poiseuille flow; and for 
banding and coexistence of isotropic and nematic phases under shear.
\end{abstract}

\begin{multicols}{2}
There is growing interest in obtaining a fundamental physical
understanding of the flow properties of liquid
crystals, polymer melts, and droplet suspensions. The hydrodynamics of
such complex fluids can be complicated and very different from
that of simple liquids because of the coupling between the microscopic
structure and the velocity fields imposed by the flow \cite{L88}. Examples
include shear-thinning and thickening and  non-equilibrium phase
transitions such as banding under shear \cite{O92,O99,C89}.

To understand the physics underlying such flow properties it
is helpful to develop simulation techniques to probe the hydrodynamics of
complex fluids. This has proved difficult because of the diverse
length and time scales involved. Microscopic approaches, such as
molecular dynamics, provide the most faithful representation of the
microscopic physics. However they are not usually able to probe
hydrodynamic time scales. 

One attempt to surmount this problem has been the development of lattice
Boltzmann simulations \cite{C98}. These solve the hydrodynamic equations
of motion while inputting sufficient, albeit generic, molecular information to
model the important physics of a given fluid. This is often
done by imposing a Landau free energy functional, so the fluid
evolves to a known thermodynamic equilibrium \cite{S96}.

Most of the results thus far using lattice Boltzmann approaches have
concentrated on the properties of binary fluid mixtures. 
Considerable progress has been made in understanding the
effect of hydrodynamics on domain growth in 2- and 3-dimensions \cite{K99}.
Investigations of flow have
been more limited although there has been some work on binary
solutions under oscillatory shear, on the flow of binary mixtures
in porous media, and on amphiphilic fluids \cite{C98}.

In this paper we describe a lattice Boltzmann algorithm for liquid
crystal hydrodynamics. The aim is to investigate the wide range of
physical phenomena which result because the director field couples to
the flow.  First we assess the effect of
hydrodynamics on defect coalescence, a process important for phase
ordering.  This illustrates how the director configuration induces
flow.  We then examine the orientation of the director in 
Poiseuille flow.  In addition to a steady state director
configuration for slower fluid velocities, the director can exhibit
tumbling and chaotic orbits in rapid flows.  Finally, we study the
system under shear and demonstrate how the liquid crystal undergoes a
non-equilibrium phase transition to a banded state where the bands are
different phases coexisting at different strains but a unique
stress \cite{O92,O99}.  

We emphasize that almost all previous work has either
ignored the flow, imposed a flow and worked out the director
configuration ignoring back-flow effects, or coupled states of
the director with states of flow using arguments based on the
stability of interfaces.  Here results are obtained self-consistently,
 by simulating the full hydrodynamics equations coupled to a
description of the liquid crystal based on a tensor order parameter. 

First we present the relevant equations of motion and the
extensions of the lattice Boltzmann approach needed to solve them. A
major difference from more simple
fluids is that liquid crystals are described by a {\em tensor} order
parameter ${\bf Q}$ (related to the director ${\vec n}$ 
by $Q_{\alpha\beta}=\langle n_\alpha n_\beta -
{1\over 3} \delta_{\alpha\beta}\rangle$). Their equilibrium properties
can be described by the Landau--de Gennes free energy functional
\cite{dG93,Doi} 
\begin{eqnarray}
{\cal F}&=&k_B T \int d^3 r {\phi}\left\{ 
\frac{1}{2}(1- \frac{\gamma}{3}) Q_{\alpha \beta} Q_{\beta \alpha} 
- \frac{\gamma}{3} Q_{\alpha \beta}Q_{\beta \zeta}Q_{\zeta \alpha}\right.\nonumber\\
& &\qquad\left.+ \frac{\gamma}{4} (Q_{\alpha \beta}Q_ {\beta \alpha})^2
+ \frac{\kappa}{2} (\partial_\alpha Q_{\beta \zeta})^2
  \right\}
\label{Qfree}
\end{eqnarray}
where $\phi$ is the liquid crystal concentration and $T$ the
temperature.  The exact form of the coefficients of the bulk free
energy terms is not important and for simplicity we write them in
terms of a single parameter $\gamma$ \cite{foot1}. 
Similarly we restrict ourselves
in the first instance to a single elastic constant $\kappa$.

The order parameter is not conserved. It evolves according to the
convection--diffusion equation \cite{O99,Doi,FC98}
\begin{equation}
(\partial_t+\partial_{\alpha} u_\alpha){\bf Q}=
{\bf S}({\bf W},{\bf Q})
+(\Gamma/k_B T \phi){\bf H}({\bf Q})
\label{cd}
\end{equation}
where ${\vec u}$ is the bulk fluid velocity and $\Gamma$ is a diffusion
constant. Of the terms on the right-hand side of Eqn.~(\ref{cd})
\begin{eqnarray}
{\bf S}({\bf W},{\bf Q})
&=&(\lambda{\bf D}+{\bf \Omega})({\bf Q}+{1\over 3}{\bf I})+({\bf Q}+{1\over 3}{\bf I})(\lambda{\bf D}-{\bf \Omega})\nonumber\\
& & -2\lambda({\bf Q}+{1\over 3}{\bf I}){\mbox{Tr}}({\bf Q}{\bf W})
\end{eqnarray}
accounts for the rotation of the nematic order parameter driven by the
symmetric ${\bf D}$, and antisymmetric ${\bf \Omega}$, parts of the
velocity gradients $W_{\alpha \beta}=\partial_\beta u_\alpha$.  The
parameter $\lambda$ is related to the aspect ratio of the polymer
molecule or alternatively can be viewed as a phenomenological
parameter to correct for inaccuracies of quadratic closure \cite{FC98}.
The molecular field, 
\begin{equation}
{\bf H}
=-\left\{ {\delta {\cal F} \over \delta {\bf Q}}-{1\over 3}{\bf
    I}{\mbox{Tr}}{\delta {\cal F} \over \delta {\bf Q}}\right\}
\end{equation}
describes the evolution to equilibrium in a way analogous to Model A.

The flow of the liquid crystal fluid obeys the continuity and
Navier-Stokes equations
\begin{equation}
\partial_t \rho + \partial_{\alpha} \rho u_{\alpha}= 0,
\label{cont}
\end{equation}
\begin{eqnarray}
& & \rho\partial_t u_\alpha+\rho u_\beta \partial_\beta
u_\alpha=\partial_\beta \tau_{\alpha\beta}+\partial_\beta
\sigma_{\alpha\beta}+\nonumber\\
& & \qquad{2 \rho \tau_f /
3}(\partial_\beta(\delta_{\alpha\beta}+3\partial_n
\sigma_{\alpha\beta})\partial_\zeta u_\zeta+\partial_\alpha
u_\beta+\partial_\beta u_\alpha)
\label{ns}
\end{eqnarray}
where 
\begin{eqnarray}
{\bf \sigma}_{\alpha \beta}&=&-\rho T \delta_{\alpha\beta}-3
{\bf H}_{\alpha\beta}-\nonumber\\
& & \qquad\kappa (\partial_\alpha Q_{\zeta\delta} \partial_\beta
Q_{\zeta\delta}- \partial_\lambda
Q_{\zeta\delta}\partial_\lambda Q_{\zeta\delta}
\delta_{\alpha\beta}/2)    \\
{\bf \tau}_{\alpha \beta}&=&{\bf H}{\bf Q}-{\bf Q}{\bf H}
\label{stress}
\end{eqnarray}
are the symmetric and antisymmetric contributions to the
non-dissipative part of the stress tensor respectively.

A lattice Boltzmann scheme which reproduces equations (\ref{Qfree}) to
(\ref{stress}) to 
second order can be defined in terms of two distribution
functions $f_i (\vec{x})$ and ${\bf G}_i (\vec{x})$ where $i$ labels
lattice directions from site $\vec{x}$. Physical variables are related to 
the distribution functions by
\begin{equation}
\rho=\sum_i f_i, \qquad \rho u_\alpha = \sum_i f_i  e_{i\alpha},
\qquad {\bf Q} = \sum_i {\bf G}_i.
\end{equation}

The distribution functions evolve in a time step $\Delta t$ according to
\begin{eqnarray}
& &f_i({\vec x}+{\vec e}_i \Delta t,t+\Delta t)-f_i({\vec x},t)=
\frac{\Delta t}{2} \left[{\cal C}_{fi}({\vec x},t,\left\{f_i \right\})+\right.\nonumber\\
& &\qquad\qquad\qquad\left. {\cal C}_{fi}({\vec x}+{\vec e}_i \Delta t,t+\Delta
t,\left\{f_i^*\right\})\right]
\nonumber\\ 
& &{\bf G}_i({\vec x}+{\vec e}_i \Delta t,t+\Delta t)-{\bf G}_i({\vec x},t)=\frac{\Delta t}{2}\left[ {\cal C}_{{\bf G}i}({\vec x},t,\left\{{\bf G}_i \right\})+\right.\nonumber\\
& &\qquad\qquad\qquad\left.
                {\cal C}_{{\bf G}i}({\vec x}+{\vec e}_i \Delta
                t,t+\Delta t,\left\{{\bf G}_i^*\right\})\right]
\label{evol}
\end{eqnarray}
where the collision operators are taken to have the form of a single
relaxation time Boltzmann equation, together with a forcing term
\begin{eqnarray}
& &{\cal C}_{fi}({\vec x},t,\left\{f_i \right\})= \nonumber\\
& &\quad-\frac{1}{\tau_f}(f_i({\vec x},t)-f_i^0({\vec x},t,\left\{f_i \right\}))
+p_i({\vec x},t,\left\{f_i \right\}),\nonumber\\ 
& &{\cal C}_{{\bf G}i}({\vec x},t,\left\{{\bf G}_i \right\})=\nonumber\\ 
& &\quad-\frac{1}{\tau_{\bf G}}({\bf G}_i({\vec x},t)-{\bf G}_i^0({\vec x},t,\left\{{\bf G}_i \right\}))
+h_i({\vec x},t,\left\{{\bf G}_i \right\}).
\label{collision}
\end{eqnarray}
$f_i^*$ and ${\bf G}_i^*$ in equations (\ref{evol}) and (\ref{collision})
are first order approximations to $f_i({\vec x}+{\vec e}_i \Delta
t,t+\Delta t)$ and ${\bf G}_i({\vec x}+{\vec e}_i \Delta t,t+\Delta
t)$. They are introduced to remove lattice viscosity terms to second
order and they give improved stability.

The form of the equations of motion and thermodynamic equilibrium
follow from the choice of the moments of the equilibrium distributions
$f^0_i$ and ${\bf G}^0_i$ and the driving terms $p_i$ and
$h_i$. $f_i^0$ is constrained by
\begin{eqnarray}
&&\sum_i f_i^{0} = \rho,\qquad \sum_i f_i^{0} e_{i \alpha} = \rho
u_{\alpha}, \nonumber\\ 
&&\sum_i f_i^{0} e_{i\alpha}e_{i\beta} = -\sigma_{\alpha\beta}+\rho
u_\alpha u_\beta
\label{mom1} 
\end{eqnarray}
where the zeroth and first moments are chosen to impose conservation of
mass and momentum. The second moment of $f^{0}$ controls the symmetric
part of the stress tensor, whereas the moments of $p_i$
\begin{eqnarray}
&&\sum_i p_i = 0, \quad \sum_i p_i e_{i\alpha} = \partial_\beta
\tau_{\alpha\beta},\quad \sum_i p_i
e_{i\alpha}e_{i\beta} = 0
\end{eqnarray}
impose the antisymmetric part of the stress tensor.

For the equilibrium of the order parameter distribution we choose
\begin{eqnarray}
 &&\sum_i {\bf G}_i^{0} = {\bf Q},\quad \sum_i
{\bf G}_i^{0} {e_{i\alpha}} = {\bf Q}{u_{\alpha}},\nonumber\\
&&\quad \sum_i {\bf G}_i^{0}
e_{i\alpha}e_{i\beta} = {\bf Q} u_\alpha u_\beta .
\end{eqnarray}
This ensures that the fluid minimizes its free energy at equilibrium
and that it is convected with the flow. Finally the evolution of the
order parameter is most conveniently modeled by choosing
\begin{eqnarray}
 &&\sum_i h_i =(D/k_B T \phi){\bf H}({\bf Q})
+{\bf S}({\bf W},{\bf Q}),\nonumber\\ 
&&\qquad \sum_i h_i {e_{i\alpha}} = (\sum_i h_i) {u_{\alpha}}.
\label{mom4}
\end{eqnarray}
Conditions (\ref{mom1})--(\ref{mom4}) can be satisfied by taking
$f_i^{0}$, ${\bf G}_i^{0}$, $h_i$, and $p_i$ as polynomial
expansions in the velocity as is usual in lattice Boltzmann
schemes\cite{C98}.
A second order Chapman--Enskog expansion for the evolution equations 
(\ref{evol}) incorporating the conditions
(\ref{mom1})--(\ref{mom4}) leads to the equations of motion
(\ref{cd}), (\ref{cont}), and (\ref{ns}).

The alternative Ericksen--Leslie--Parodi (ELP) equations of liquid crystal 
hydrodynamics are written in terms of the evolution of the director
field ${\bf n}$ \cite{dG93} rather than the tensor order
parameter ${\bf Q}$.  We use the latter approach for two reasons.
First, the motion of disclination lines (points in two dimensions) is 
explicitly included and it is particularly interesting to assess the
effect of flow on this dynamics. Second, we are interested in examining
phase transitions induced by shear and so need to describe
both the isotropic and nematic phases.  The ELP formalism describes
only the nematic phase at a fixed amplitude of order parameter.  If we
restrict the order parameter to be uniaxial with fixed
amplitude, the ${\bf Q}$ equations of motion can be reduced to the ELP
equations.  However, due to our choice of coefficients in, for
example the free energy, we have specialized to three independent
viscosity coefficients rather than the five in the ELP formalism.
A lattice Boltzmann method for implementing the ELP equations is being
studied elsewhere \cite{Care}

\begin{figure}
\narrowtext 
\centerline{\epsfxsize=3.2in 
\epsffile{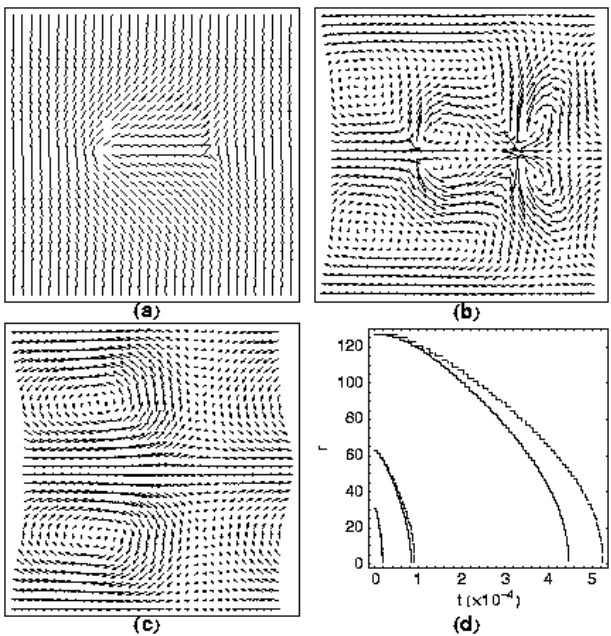}} 
\vskip 0.4true cm
\caption{(a) Director field for $\pm 1/2$ disclinations moving towards
  annihilation.  (b) Fluid velocity induced
  by the disclinations at the same time as the configuration shown in
  (a).  (c) Vortices left in the fluid velocity field after the
  defects have annihilated.  (d) Separation of the disclinations as a
  function of time for systems of sizes $L=64$, $128$ and $256$ (left
  to right) with (solid) and without (dashed) hydrodynamics.  The
  disclinations start out at rest at a separation of $L/2-1$.  
  Defects in systems with hydrodynamics annihilate faster.}
\label{anhil}
\end{figure}

If Fig.~\ref{anhil} we show the flow fields induced when two disclinations of
opposite sign mutually annihilate, a process important in
phase separation.  The director field induces a flow in the fluid in
the form of vortices accompanying the disclinations \cite{foot3}.   
Even after the defects have annihilated, they leave vortices behind in the
fluid which then gradually decay away.  The vortices induced in the
fluid speed up the annihilation process, but do not change the $t^{1/2}$
power law \cite{Z95,CD96} for annihilation.  This is because the vortex-vortex
interaction is of the same form as the disclination-disclination
interaction.  However, in a system with many disclinations the
interaction between annihilating pairs is likely to be
much stronger when hydrodynamics is present as the fluid vortices left by the
annihilation process are at a considerable distance from the
annihilation site.  We are currently examining the effect of
hydrodynamics on the phase ordering.

\begin{figure}
\centerline{\epsfxsize=3.2in 
\epsffile{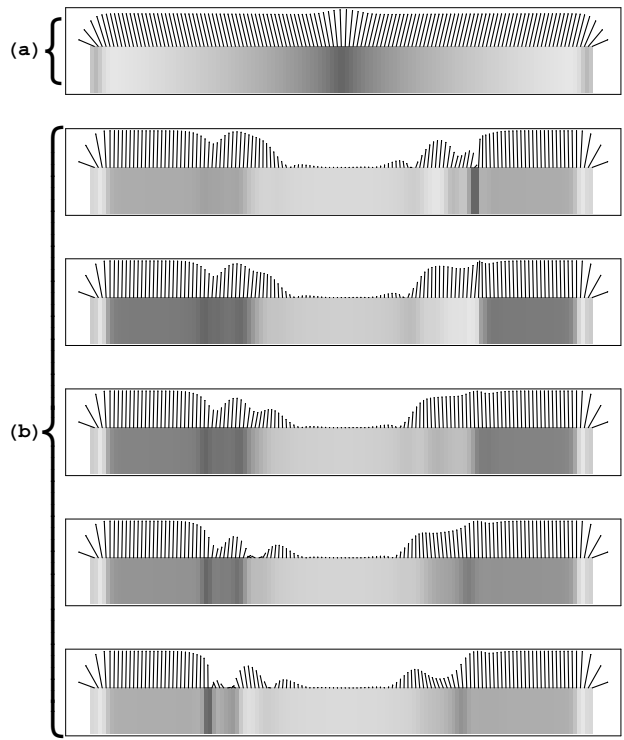}} 
\vskip 0.4true cm
\caption{Two different states in Poiseuille flow,
  where the lines represent the director orientation (eigenvector
  corresponding to largest eigenvalue of ${\bf Q}$) projected down onto the
  $xy$-plane, and shading represents the amplitude of the order
  parameter (largest eigenvalue).  Flow is from top to bottom, and the
  walls are at the left and right. At the walls, the director is aligned
  perpendicular to the boundary.  (a) A stable configuration
  at low flow.  (b) Snapshots of an oscillating configuration  
  where the central region is in the ``log-rolling state'' (director
  perpendicular to the plane) and the boundary region consists of a
  transition from a configuration in the shear plane to a 
  ``tumbling'' and ``kayaking'' region (director rotating in and out of
  the plane) interfacing to the central log-rolling state.}
\label{pois}
\end{figure}

A second example of the coupling between the director field and the
flow is shown in Fig.\ref{pois} where a Poiseuille flow is imposed on the
liquid crystal with the director field pinned perpendicular to the flow
direction at the boundaries. The director field couples to the shear
component of the flow leading to a rich variety of possible dynamical
states.  For slow flows a static director field configuration is
induced by the flow (Fig.~\ref{pois}(a)).  We note that in order to
match the director fields at the midpoint of the flow, the director
orients {\it parallel} to the flow, not perpendicular as has been
previously suggested \cite{dG93}, as this configuration has the lower elastic
energy\cite{foot2}.  For faster, more rotational flows, or
alternatively smaller $\lambda$, the ``log rolling'' state, where the
director points perpendicular to the shear plane, has a lower
viscosity and is more stable in the bulk.  However, as the director is
pinned in the shear plane at the boundary it must match the
boundary configuration onto the log rolling state in the interior.
The system is unable to do this with a steady state configuration and
the director tumbles and kayaks (rotates in and out of the plane) in
the interface region, as show in Figure~\ref{pois}(b).  
While the velocity profile remains nearly parabolic, as would be
expected for Poiseuille flow in a simple fluid, the effective
viscosity is strongly influenced by the state of the director.

\begin{figure}
\narrowtext 
\centerline{\epsfxsize=3.2in 
\epsffile{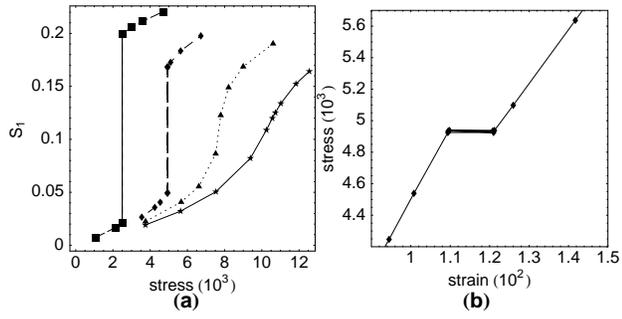}} 
\vskip 0.4true cm
\caption{(a) Amplitude of the order parameter $S_1$(largest eigenvalue of
  ${\bf Q}$) as a function of the shear stress for $\gamma=2.65$
  (squares), $2.6$ (diamonds), $2.55$ (triangles) and $2.5$ (stars).
  The equilibrium transition occurs at $\gamma\approx2.7$. (b) The
  stress-strain curve shows a discontinuity at the point where the
  shear stress induces a first-order, non-equilibrium phase transition
  from the weakly ordered paranematic I phase to the strongly ordered
  nematic N phase.}
\label{shrbnd}
\end{figure}

Shear acts to orient the director field with respect to the flow and
in this sense is analogous to a magnetic field in spin systems.
As such, it can shift the transition from the isotropic (I) to nematic
(N) phases\cite{O92,O99,S90}.  The resulting non-equilibrium phase
diagram in the shear-stress--$(1-\gamma/3)$ plane (see
Eqn.~\ref{Qfree}) consists of a line of first order phase transitions
connecting to the equilibrium transition at zero stress and
terminating at a non-equilibrium critical point.  This is illustrated
in Figure~\ref{shrbnd}(a) which shows the behavior of $S_1$, the largest 
eigenvalue of ${\bf Q}$, as a function of the shear stress.  
As $\gamma$ is increased, the jump in the
order parameter decreases and disappears at a non-equilibrium critical point.

In an experiment, or in our simulations, we do not have direct control over
the shear stress (i.e. the total off diagonal stress, including
elastic and viscous terms), but rather control the average strain
rate (the relative velocity of the walls).  The coexistence region is
just a line on the stress-temperature phase diagram, but it
corresponds to a finite region on a strain-temperature diagram.  In
fact the two different coexisting phases have a {\it different} strain
rate.  This phenomenon is known as shear banding and  results in a
plateau in the stress-strain curve shown in Figure~\ref{shrbnd}(b).
Once coexistence is reached, increasing the strain rate does not
increase the stress, it just converts more of the system from 
the I to N phase.  Once all the system is in the N phase, the stress
again starts to increase.  Our results are in agreement with those
obtained in Ref.\cite{O92,O99}.  There an interface was imposed on
a one-dimensional system and its stability was used to infer the
state of the system.  Here, shear banding is spontaneous and work is
in progress to investigate the way in which the bands form.

In conclusion, we have derived and implemented a lattice Boltzmann
algorithm for liquid crystal hydrodynamics.  This opens the way to
investigate a wide range of physical phenomena which result from the
coupling between the director field and the flow, many of which have
been examined only indirectly or with severe approximations
in the past. 

We thank G. Gonnella, C. Care and P. Olmsted for useful conversations.  
This work was funded in part by grants from the EPSRC.

\vfill

\vfill

\vfill

\vfill

\end{multicols}

\begin{references}
\bibitem{foot1}
$\gamma$ is the Doi excluded volume parameter.
\bibitem{foot2}
The director aligns with an angle $\theta$ to the flow given by $\cos
2\theta=1/\xi$ where $\xi$ is related to coefficients in
equation (\ref{cd}).  Since this implies $\theta < \pi/4$  it will
always cost less energy to have $\theta=0$ at the center rather than
$\theta=\pi/2$ as suggested in \cite{dG93}, pg 219.
\end{references}
\end{document}